\documentclass{jfm}
\usepackage{graphicx}
\usepackage{dcolumn}
\usepackage{bm}
\usepackage{color}
\usepackage{placeins}
\usepackage{array}
\usepackage{subfigure}
\usepackage{natbib}
\usepackage{amssymb}
\usepackage{amsbsy}

\title[Radial boundary layer structure and Nusselt number in Rayleigh-B\'enard convection]
{Radial boundary layer structure and Nusselt number in Rayleigh-B\'enard convection}

\author[Richard J.A.M. Stevens, Roberto Verzicco and Detlef Lohse]
{Richard J.A.M. Stevens$^1$, Roberto Verzicco$^2$ and Detlef Lohse$^1$}

\affiliation{
$^1$Department of Science and Technology and J.M. Burgers Center for Fluid Dynamics, University of Twente, P.O Box 217, 7500 AE Enschede, The Netherlands,\\
$^2$Dept. of Mech. Eng., Universita' di Roma "Tor Vergata",Via del Politecnico 1, 00133, Roma.}

\pubyear{1996}
\volume{538}
\pagerange{119--126}
\date{?? and in revised form ??}

\begin{document}

\maketitle

\begin{abstract}
Results from direct numerical simulations for three dimensional Rayleigh-B\'enard convection in a cylindrical cell of aspect ratio $1/2$ and $Pr=0.7$ are presented. They span five decades of $Ra$ from $2\times10^6$ to $2\times10^{11}$. Good numerical resolution with grid spacing $\sim$ Kolmogorov scale turns out to be crucial to accurately calculate the Nusselt number, which is in good agreement with the experimental data by Niemela \emph{et al.}, Nature, \textbf{404}, 837 (2000). In underresolved simulations the hot (cold) plumes travel further from the bottom (top) plate than in the fully resolved case, because the thermal dissipation close to the sidewall (where the grid cells are largest) is insufficient. We compared the fully resolved thermal boundary layer profile with the Prandtl-Blasius profile. We find that the boundary layer profile is closer to the Prandtl Blasius profile at the cylinder axis than close to the sidewall, due to rising plumes in that region.
\end{abstract}

\section{Introduction}

Turbulent Rayleigh-B\'enard convection (RBC), continues to be a topic of intense research (\cite{ahl09,loh10}). The system is relevant to numerous astro- and geo-physical phenomena, including convection in the arctic ocean, in Earth's outer core, in the interior of gaseous giant planets, and in the outer layer of the Sun. Thus the problem is of interest in a wide range of sciences, including geology, oceanography, climatology, and astrophysics.

For given aspect ratio $\Gamma \equiv D/L $ ($D$ is the cell diameter and $L$ its height) and given geometry, the nature of RBC is determined by the Rayleigh number $Ra = \beta g \Delta L^3/(\kappa\nu)$ and by the Prandtl number $Pr = \nu /\kappa$. Here, $\beta$ is the thermal expansion coefficient, $g$ the gravitational acceleration, $\Delta = T_b-T_t$ the difference between the imposed temperatures $T_b$ and $T_t$ at the bottom and the top of the sample, respectively, and $\nu$ and $\kappa$ the kinematic viscosity and the thermal diffusivity, respectively. When $Pr$ is fixed, $Ra$ is the only control parameter. There is still no universally accepted theory what the asymptotic high Rayleigh number Nu(Ra) relationship should be (\cite{kra62,spi71,cas89,shr90,ahl09}), and the experimental results are controversial (\cite{hes87,cha97,roc02,nie00,nie01,nie03,nik05,fun05,fun09}).

\begin{figure}
\centering
\subfigure[]{\includegraphics[height=0.35\textwidth]{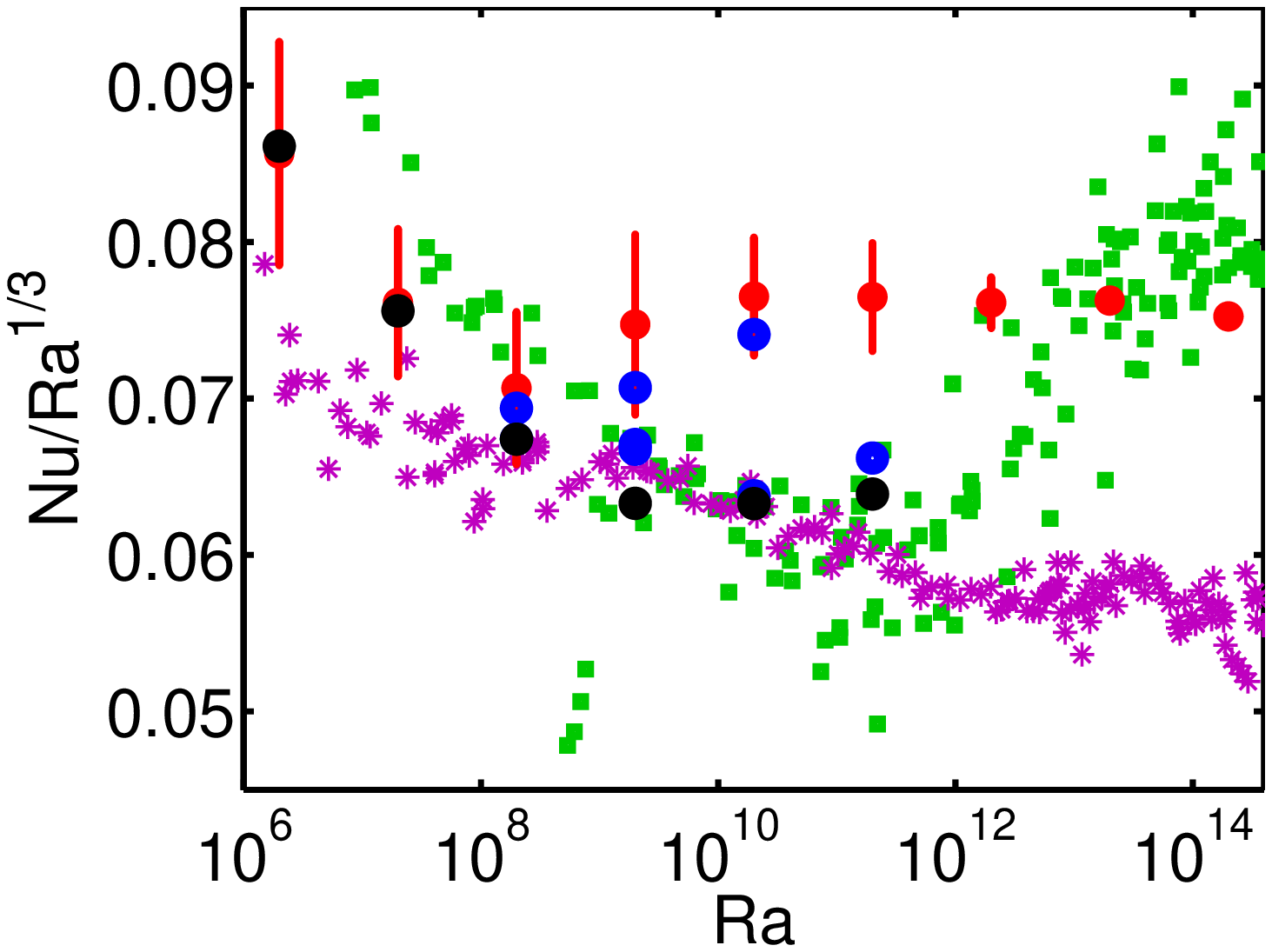}}
\hspace{5mm}
\subfigure[]{\includegraphics[height=0.35\textwidth]{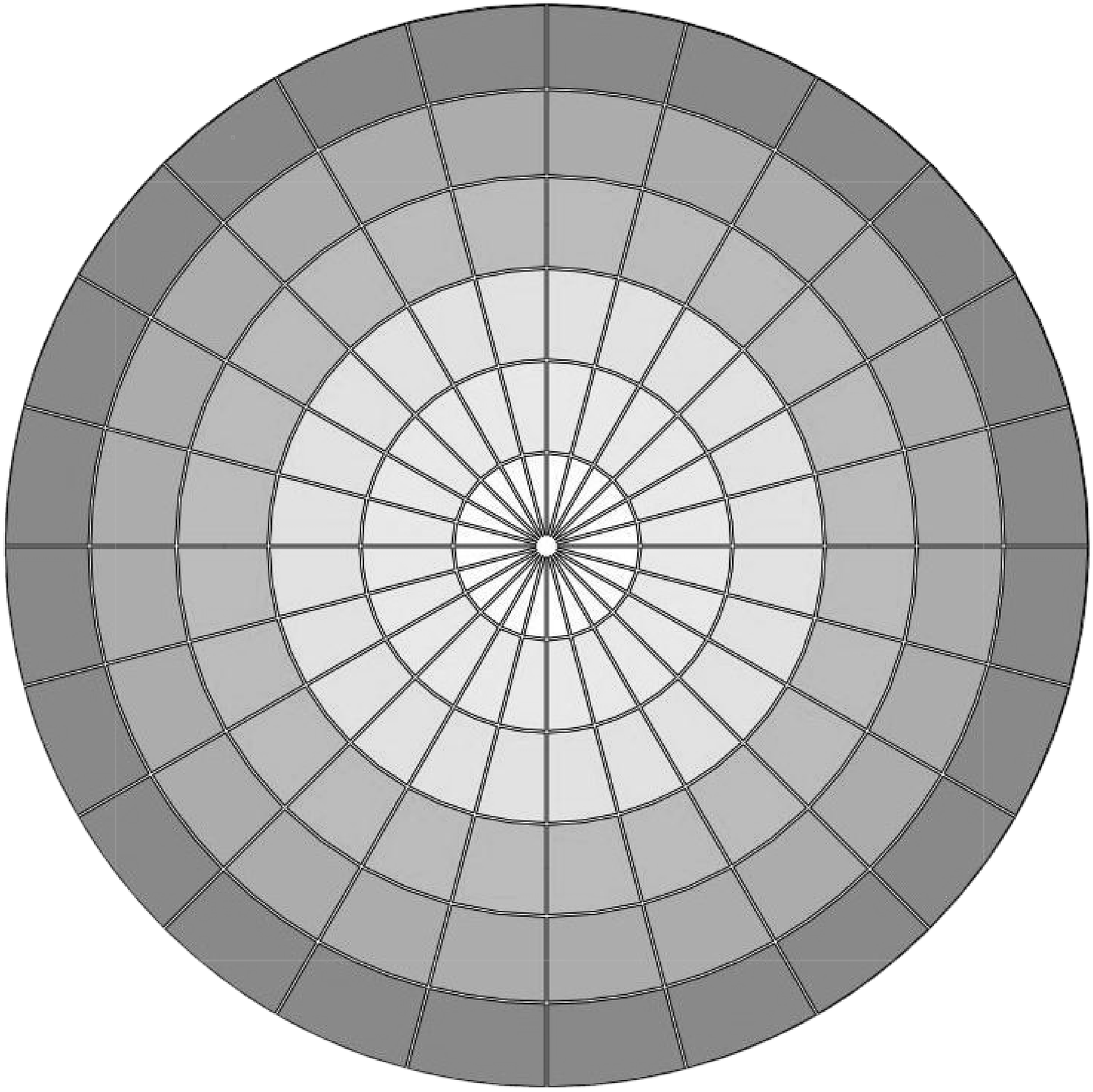}}
\caption{(a) Compensated Nusselt number vs the Rayleigh number for $Pr=0.7$. Purple stars are experimental data from \cite{nie00} and the green squares are from \cite{cha01}. The DNS results from \cite{ver03} and \cite{ama05} are indicated in red and the present DNS results with the highest resolution are indicated by the black dots (the error is smaller than the dot size). The results of the underresolved simulations of this study are indicated by the blue dots. (b) Sketch of the grid geometry. The cells close to the sidewall are largest and therefore this region is least resolved.}
\label{fig:Nusselt}
\end{figure}

For more moderate $Ra$ up to $2\times 10^{14}$ previous direct numerical simulations (DNS) by \cite{ama05} in a three dimensional cylindrical cell of aspect ratio $1/2$ with $Pr =0.7$ showed a higher Nusselt $Nu$ number than measured in experiments, see figure \ref{fig:Nusselt}. To explain this discrepancy it was then suggested by \cite{ver08} that the experimental conditions are closer to fixed flux conditions than fixed temperature conditions. However, recent two dimensional simulations by \cite{joh09} showed that $Nu$ obtained in simulations with constant temperature and constant heat flux are identical when $Ra \gtrsim 5 \times 10^6$. In this paper we show that the Nusselt number obtained in the three dimensional simulations with constant temperature conditions is in good agreement with the experimental data, see figure \ref{fig:Nusselt}, when the resolution is sufficiently high.

\section{Numerical method and results on the Nusselt number}

We numerically solved the three-dimensional Navier-Stokes equations within the Boussinesq approximation,
\begin{eqnarray}
 \frac{D\textbf{u}}{Dt} &=& - \nabla P + \left( \frac{Pr}{Ra} \right)^{1/2} \nabla^2 \textbf{u} + \theta \textbf{$\widehat{z}$}, \\
 \frac{D\theta}{Dt} &=& \frac{1}{(PrRa)^{1/2}}\nabla^2 \theta ,
\end{eqnarray}
with  $\nabla \cdot \textbf{u} = 0$.  Here \textbf{$\widehat{z}$} is the unit vector pointing in the opposite direction to gravity, $D/Dt = \partial_t + \textbf{u} \cdot \nabla $ the material derivative,  $\textbf{u}$ the velocity vector (with no-slip boundary conditions at all walls), and $\theta$ the non-dimensional temperature, $0\leq \theta \leq 1$. The equations have been made non-dimensional by using, next to $L$ and $\Delta$, the free-fall velocity $U=\sqrt{\beta g \Delta L}$. The numerical scheme is described in detail in \cite{ver96} and \cite{ver99,ver03}.

An important criterion in DNS simulations is that all the relevant flow scales, i.e. the Kolmogorov length $\eta$ and the Batchelor length $\eta_T$, are properly resolved. Since $Pr<1$, the smallest of these is $\eta$ and we determined $\eta$ by $\eta/L \approx \pi (Pr^2/RaNu)^{1/4}$ (\cite{gro83,ver03}). We used different grids to test the influence of the grid scales. In table \ref{Table Simulation parameters} the largest grid scale $\ell_{max}$ is compared to the Kolmogorov scale $\eta$ for each simulation. Note that their ratio is based on the global criterion, assuming a uniform distribution of the dissipation rates, in contrast to the observed peaking of the dissipation rates close to the (side)wall (see figure  \ref{fig:Dissipations}). This means that the resolution in the bulk is better than the indicated value, however worse close to the sidewall. The grid density near the plates has been enhanced to keep a sufficient number of nodes in the thermal boundary layer (BL) where the dissipation rates are high, see the column "$\lambda_{\theta}$" in table \ref{Table Simulation parameters}. We calculate $Nu$ as volume average and also by using the temperature gradients at the bottom and top plate. The volume average is calculated from the definition of the Nusselt number
 $Nu = (\langle u_z \theta \rangle_A -\kappa \partial_3 \langle \theta \rangle_A$)/ $\kappa \Delta L^{-1}$ (\cite{ver99}).
In addition, we average over the entire volume and time. The averages of the three methods, i.e. the volume average and the averages based on the temperature gradients at the bottom and top plate, are determined over at least $400$ dimensionless time units. The value $Nu$ in table \ref{Table Simulation parameters} gives the average value of these three. We also determined $Nu$ over the last half of our simulations, see the column "$Nu_h$" in table \ref{Table Simulation parameters}. These values are within $1\%$ of the value determined over the whole simulation, showing that our results are well converged. The maximum difference in $Nu$ obtained from the three methods, i.e. volume average and using the temperature gradients at the plates, is given in the column "conv" in table \ref{Table Simulation parameters}. We simulated $200$ dimensionless time units before we started to collect data to be sure to have reached the statistically stationary state. This is confirmed by comparing the statistics of the last part of the simulation used for the initialization and the part of the simulation used for the actual data collecting. In simulations where the field obtained at a lower $Ra$ (or a new random field) is used as initial condition, we observe a small overshoot in $Nu$, before it settles to it statistically stationary value, The long initialization runs are thus used to eliminate this effect. This is double checked by the convergence of the three different methods to calculate $Nu$. We never noticed a (significant) drift in $Nu$ when the results of the three methods are within $1\%$ of each other. For the four most demanding simulations, i.e. the bottom four cases in table \ref{Table Simulation parameters}, the criteria for time averaging had to be relaxed due to the limited CPU time available. Therefore we averaged these cases for $100$ dimensionless time units ($300$ time units for the simulation at $Ra=2\times 10^{10}$ on the $385 \times 257 \times 1025$ grid). We note that the error bars given in figure \ref{fig:Nusselt} show the error in Nusselt obtained on that particular grid. The effect of the grid resolution on the Nusselt results for the highest $Ra$ numbers is not estimated.

Since most simulations are started from a interpolated field obtained at a lower $Ra$, we recomputed $Nu$ for $Ra =2\times 10^9$ on the $193 \times 65 \times 257$ grid with a new flow field to rule out the effect of hysteresis on the obtained Nusselt results. The result is shown in \emph{italics} in table \ref{Table Simulation parameters} and is in excellent agreement with the original result. This confirms that our initialization and data collecting runs are long enough to eliminate the hysteresis effect on the Nusselt results. The simulations at $Ra=2\times 10^{10}$ have different initial conditions, i.e. different flow fields obtained at lower $Ra$ are used as initial condition, and here we also observe a good agreement considering the different grids that are used. Figure \ref{fig:Nusselt} shows that the DNS data converge to the experimental data when the resolution is increased. To obtain accurate results for the Nusselt number in DNS simulations the grid spacing has to $\leq \eta$ in the whole domain.  We note that in simulations for $Pr>1$ it is important to properly resolve the Batchelor scale $\eta_T$ over the whole domain, since $\eta_T<\eta$ when $Pr>1$ .

\begin{table}
  \centering
  \caption{The columns from left to right indicate $Ra$; the number of grid points in the azimuthal, radial, and axial direction ($N_{\theta} \times N_r \times N_z$), the Nusselt number ($Nu$) obtained after averaging the results of the three methods (see text) using the whole simulation length, the Nusselt number ($Nu_h$) after averaging the results of the three methods using the last half of the simulation, the maximum difference between the three methods ($Conv$), the number ($N_{BL}$) of points in the thermal BL, the maximum grid scale compared to the Kolmogorov scale estimated by the global criterion ($\ell_{max}$/$\eta$). The last two columns give the Nusselt number derived from the volume averaged kinetic $\langle\epsilon_u \rangle$ and thermal $\langle\epsilon_\theta \rangle$ dissipation rates compared to $Nu$ indicated in column three. The \emph{italic} line indicates a simulation started with a new flow field.}\label{Table Simulation parameters}
\begin{tabular}{|c|c|c|c|c|c|c|c|c|c|}
  \hline
  $Ra$ & $N_{\theta} \times N_r \times N_z$ & $Nu$ & $Nu_{h}$ & $Conv$ & $N_{BL}$ & $\ell_{max}$/$\eta$ & $\frac{\frac{\langle \epsilon_u \rangle}{\nu^3RaPr^{-2}/L^4}+1}{Nu}$ & $\frac{\frac{\langle \epsilon_\theta \rangle}{\kappa\Delta^2/L^2}}{Nu}$\\
  \hline
  $2 \times 10^6$          & $97 \times 49 \times 129$           & 10.85          & 10.92        & 0.32 $\%$         &  18      & 0.42       & --    & -- \\
  $2 \times 10^7$          & $129 \times 49 \times 193$          & 20.52          & 20.56        & 0.36 $\%$         &  17      & 0.66       & --    & -- \\
  $2 \times 10^8$          & $97 \times 49 \times 193$           & 40.57          & 40.71        & 0.02 $\%$         &  10      & 1.84       & 1.0062 & 0.8528   \\
  $2 \times 10^8$          & $193 \times 65 \times 257$          & 39.42          & 39.52        & 0.02 $\%$         &  13      & 0.92       & 0.9901& 0.9001 \\
  $2 \times 10^8$          & $257 \times 97 \times 385$          & 39.41          & 39.10        & 0.79 $\%$         &  19      & 0.70       & 0.9924& 0.9411 \\
  $2 \times 10^9$          & $129 \times 65 \times 257$          & 89.07          & 88.25        & 0.02 $\%$         &  6       & 3.01       & 1.0000& 0.7317 \\
  \emph{2 $\times$ 10$^9$} & \emph{193 $\times$ 65 $\times$ 257} & \emph{84.49}   & \emph{84.46} & \emph{0.45 $\%$}  &  \emph{7}& \emph{1.99}& \emph{1.0009}& \emph{0.7638}\\
  $2 \times 10^9$          & $193 \times 65 \times 257$          & 84.10          & 83.66        & 0.51 $\%$         &  7       & 1.98       & 0.9993& 0.7625 \\
  $2 \times 10^9$          & $385 \times 97 \times 385$          & 79.75          & 78.70        & 0.70 $\%$         &  10      & 1.15       & 0.9974& 0.8693 \\
  $2 \times 10^{10}$       & $129 \times 97 \times 385$          & 201.08         & 201.21       & 1.01 $\%$         & 12       & 6.56       & 1.0058& 0.8661 \\
  $2 \times 10^{10}$       & $513 \times 129 \times 513$         & 171.79         & 169.58       & 2.09 $\%$         & 19       & 1.59       & 0.9981 & 0.9234\\
  $2 \times 10^{10}$       & $385 \times 257 \times 1025$        & 173.13         & 173.30       & 0.98 $\%$         & 29       & 2.12       & -- & -- \\
  $2 \times 10^{11}$       & $769 \times 193 \times 769$         & 387.07         & 387.53       & 2.18 $\%$         & 16       & 2.31       & -- & --\\
  $2 \times 10^{11}$       & $769 \times 257 \times 1025$        & 373.64         & 368.88       & 2.03 $\%$         & 18       & 2.28   & 0.9828  & 0.8910\\
  \hline
\end{tabular}
\end{table}

\section{Dissipation rates, temperature distribution functions, and boundary layers}
\begin{figure}
\centering
\includegraphics[width=0.90\textwidth]{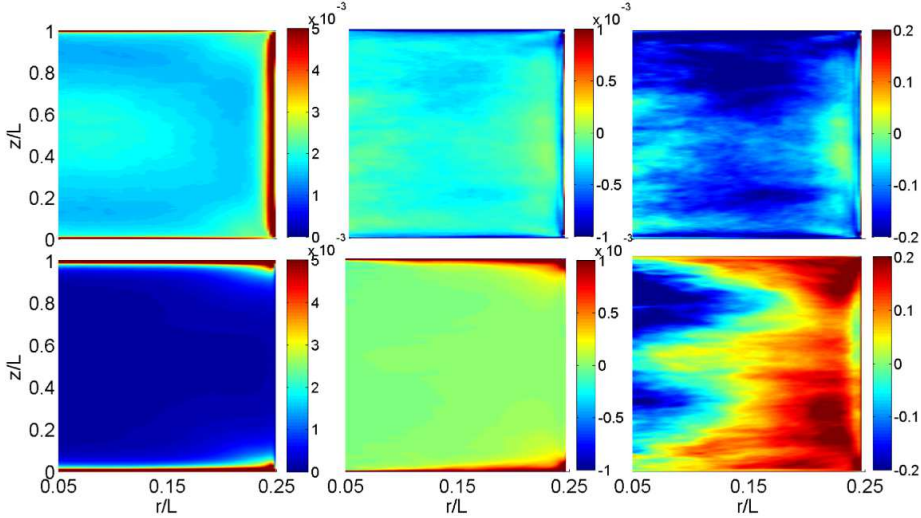}
\caption{Dimensionless kinetic (upper) and thermal (lower) dissipation rates at $Ra=2\times 10^9$. The upper row gives $\widetilde{\epsilon}_u=\epsilon_u L^3/U^2$ and the lower row $\widetilde{\epsilon}_\theta=\epsilon_\theta U/(\Delta^2L)$. The left column indicates the dimensionless kinetic $\widetilde{\epsilon}_u$ and thermal $\widetilde{\epsilon}_\theta$ dissipation rates for the high resolution case ($385 \times 97 \times 385$). The middle column gives $\widetilde{\epsilon}_u^H-\widetilde{\epsilon}_u^L$ (upper plot), and $\widetilde{\epsilon}_\theta^H-\widetilde{\epsilon}_\theta^L$ (lower plot), where the superscripts $H$ and $L$, respectively, mean the data obtained from the high ($385 \times 97 \times 385$) and low resolution simulations ($129 \times 65 \times 257$). The rightmost column gives $(\widetilde{\epsilon}_u^H-\widetilde{\epsilon}_u^L)/\widetilde{\epsilon}_u^H$ (upper plot) and $(\widetilde{\epsilon}_\theta^H-\widetilde{\epsilon}_\theta^L)/\widetilde{\epsilon}_\theta^H$ (lower plot). The difference for the thermal dissipation rates between the fully resolved and the underresolved simulations is largest (in absolute values) close to the sidewall.}
\label{fig:Dissipations}
\end{figure}

Another way to calculate $Nu$ is to look at the two exact global relations for the volume averaged kinetic and thermal energy dissipation rates $\langle \epsilon_u \rangle=\nu^3(Nu-1)RaPr^{-2}/L^4$, and $\langle \epsilon_\theta \rangle= \kappa\Delta^2 Nu /L^2$, respectively (\cite{shr90}). We have calculated the azimuthally and time averaged energy dissipation rate $\epsilon_u(\overrightarrow{x})=\nu |\nabla{\textbf{u}}|^2$ and the thermal dissipation rate $\epsilon_\theta(\overrightarrow{x})=\kappa |\nabla{\theta}|^2$. Figure \ref{fig:Dissipations} compares the difference between the dissipation rates obtained in the fully resolved and the underresolved simulations and reveals a higher thermal dissipation rate for the fully resolved simulations as it is calculated from the (temperature) gradients. In the underresolved simulations the gradients are smeared out and therefore $\epsilon_u$ and $\epsilon_\theta$ are underestimated. To check the resolution, we calculated $\epsilon_u$ and $\epsilon_\theta$ from the respective gradients and compared it with the values obtained from above global exact relations. Table \ref{Table Simulation parameters} shows that for $\epsilon_u$ the relation is basically satisfied, whereas for $\epsilon_\theta$ the difference is considerable when the simulation is underresolved, and even for the best resolved cases there is $6-8 \%$ too little dissipation. Testing above exact relations seems to be the best way to verify the grid resolution.

The vertical heat flux concentrates in the plume-dominated sidewall region where the vertical velocity reaches its maximum (\cite{sha08}). Therefore it is very important to properly resolve the region close to the sidewall. However, figure \ref{fig:Dissipations} reveals that in the underresolved simulations the region close to the sidewall is least resolved (red areas in the plot where the thermal dissipation rates are compared (right plot)), as there the grid boxes are largest, due to the cylindrical geometry of the grid (see figure \ref{fig:Nusselt}b). When the resolution is insufficient close to the sidewall, the plumes in this region, important for the heat transfer, are not properly resolved and not sufficiently dissipated. Therefore too much heat reaches the other side and correspondingly $Nu$ is overestimated in these underresolved simulations. Supplementary movies reveal the dynamics of the system for the different grid resolutions. Movie $1$ shows the temperature field close to bottom plate and movie $2$ the temperature field at mid height. Note that the smoothness of the underresolved simulations is insufficient to capture all characteristics of the flow represented in the high resolution simulation.

\begin{figure}
\centering
\subfigure[]{\includegraphics[height=0.23\textheight]{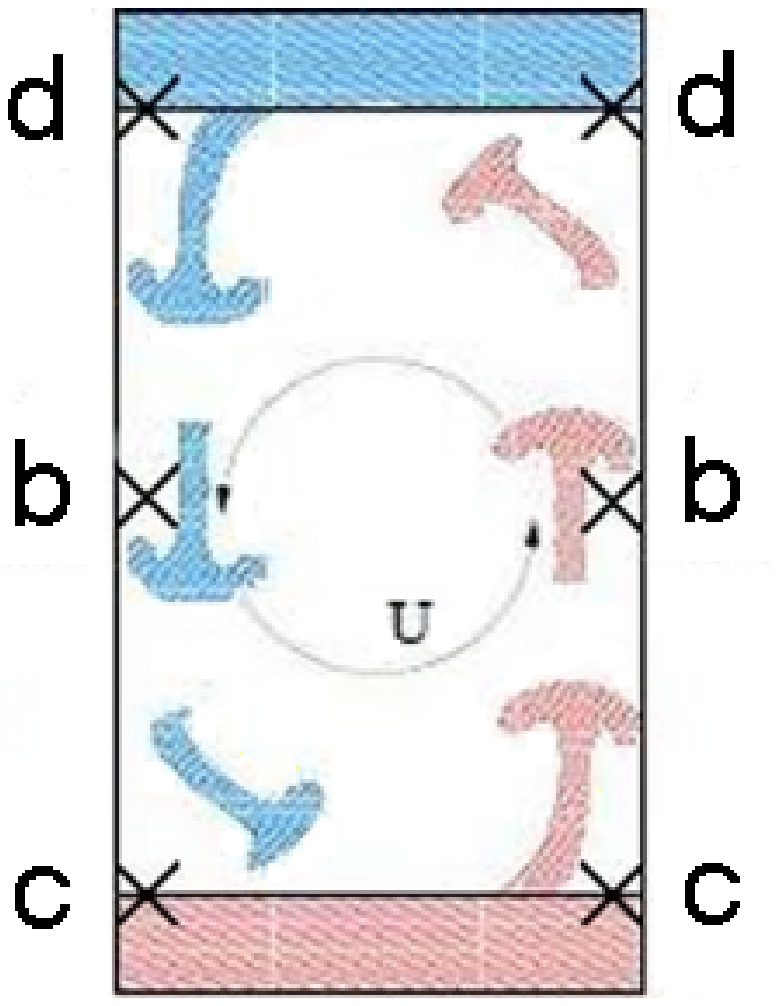}}
\subfigure[]{\includegraphics[height=0.23\textheight]{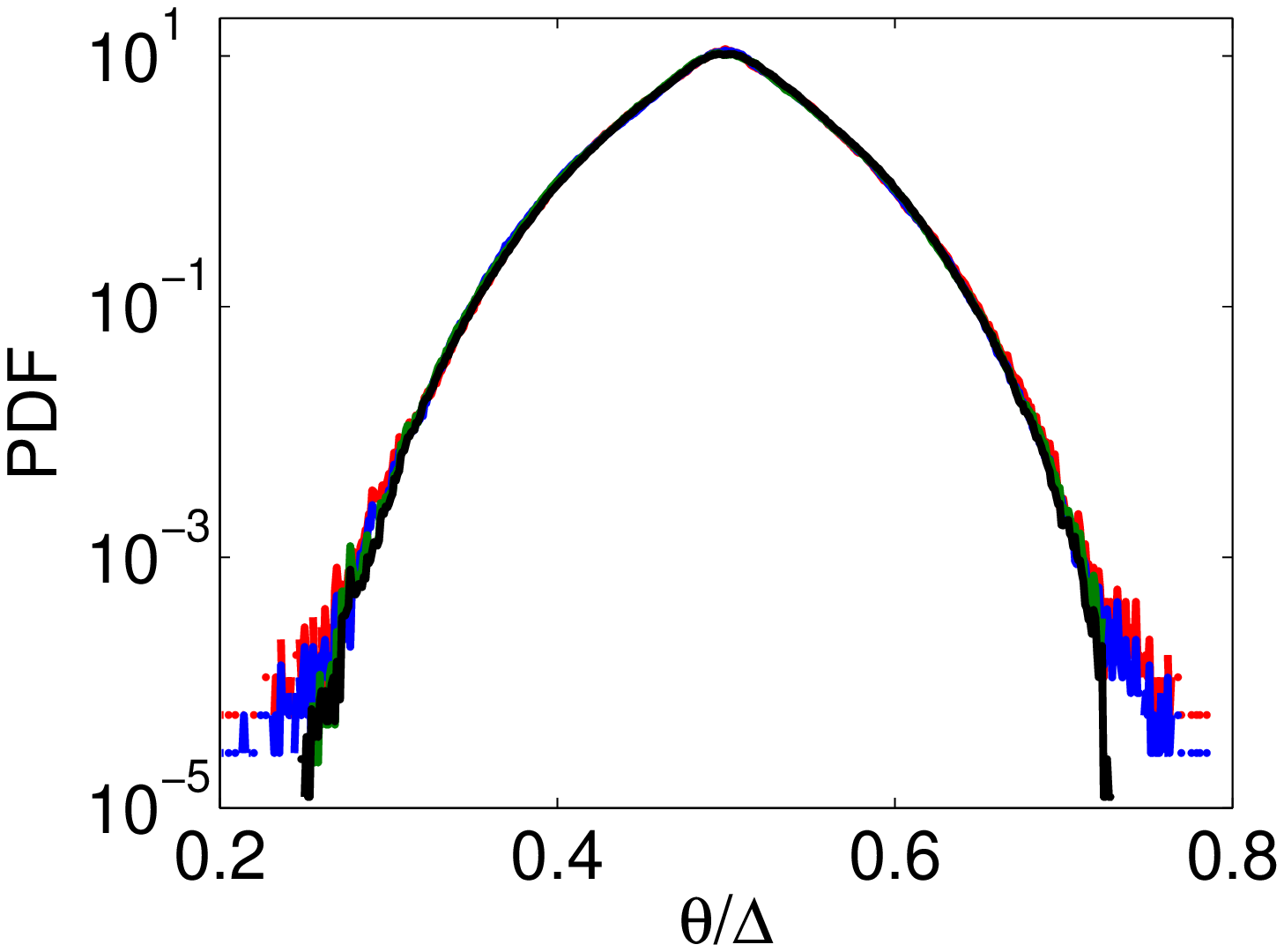}}
\subfigure[]{\includegraphics[height=0.23\textheight]{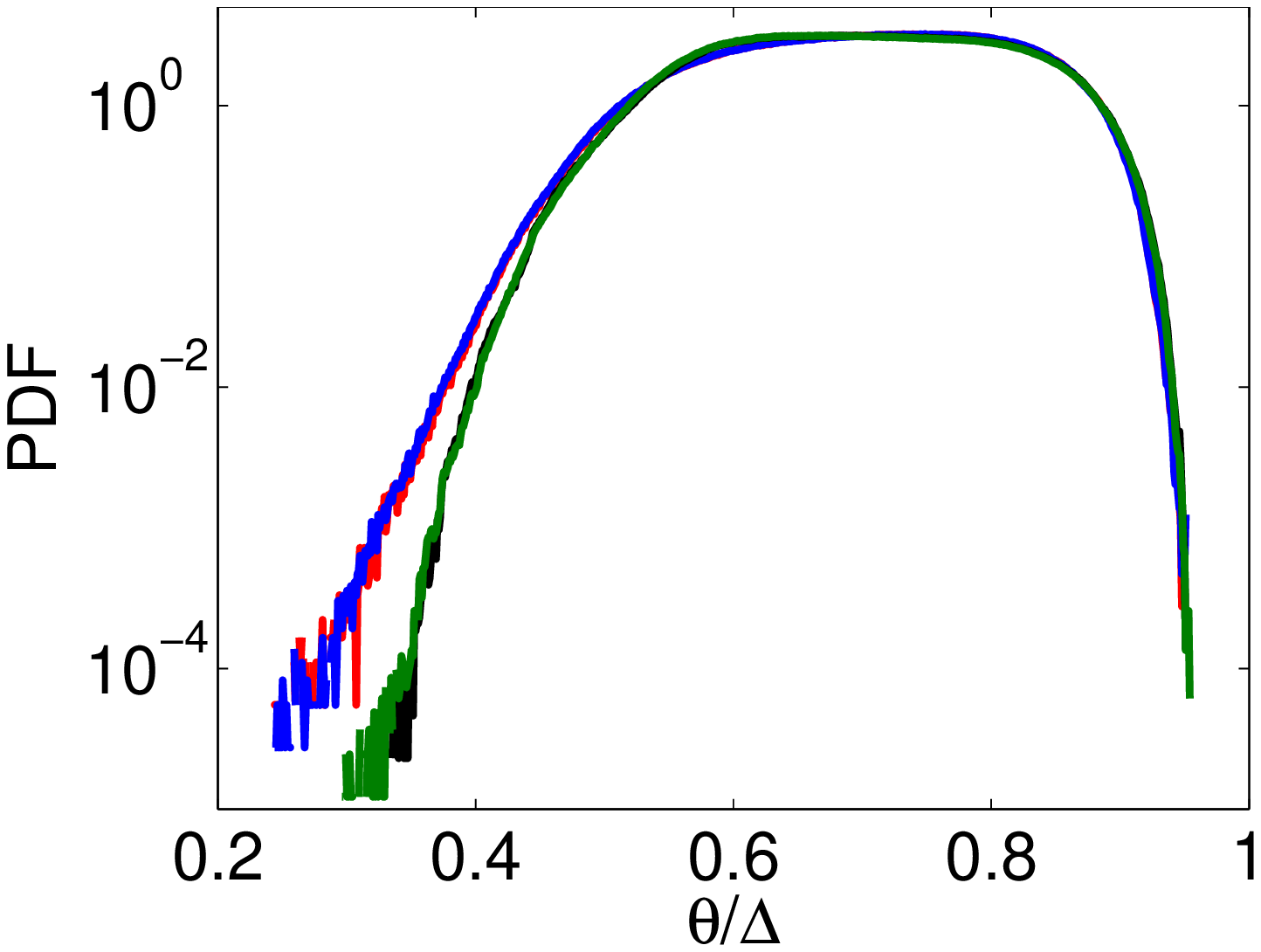}}
\subfigure[]{\includegraphics[height=0.23\textheight]{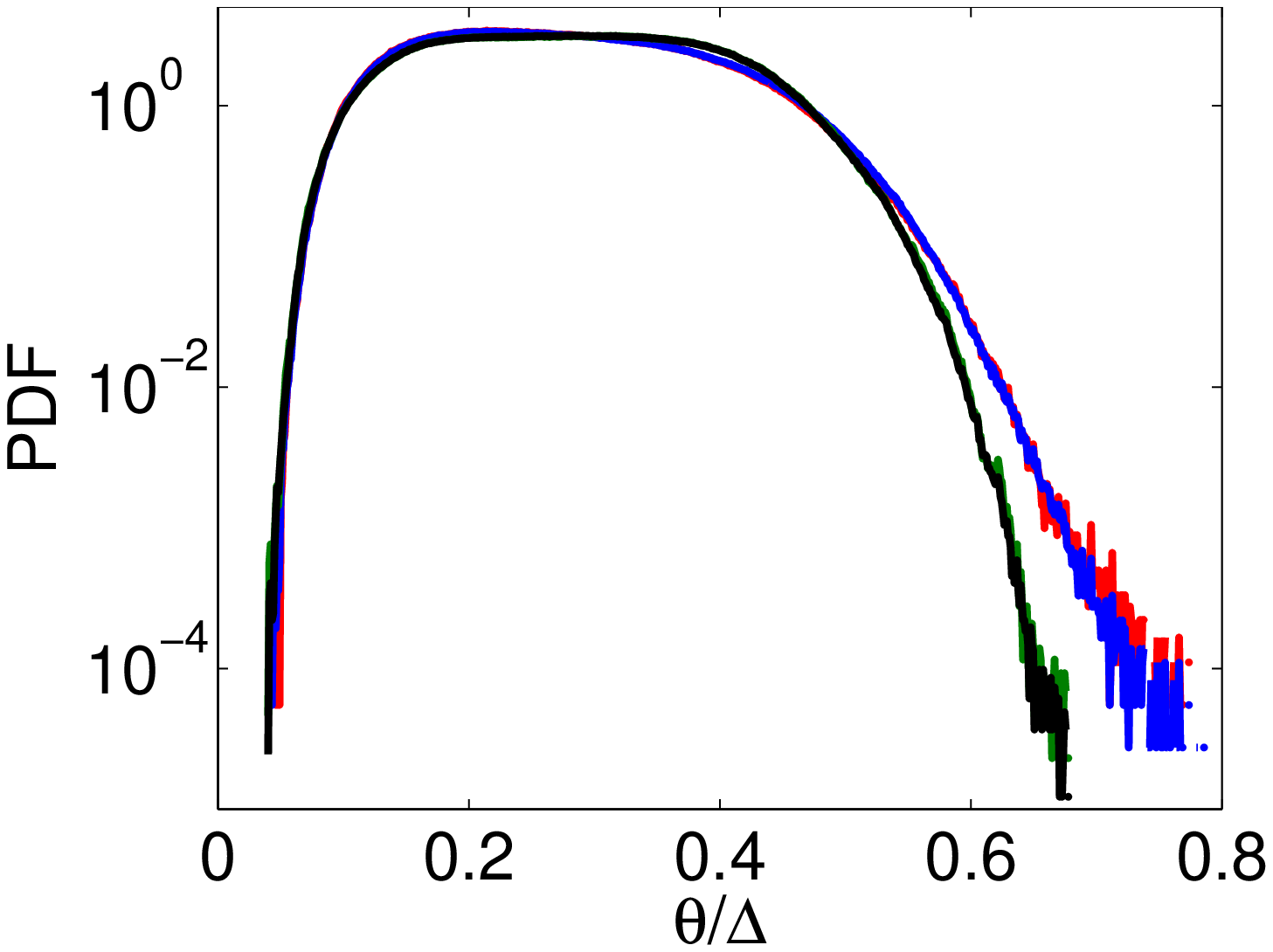}}
\caption{a) A sketch showing the locations (crosses) of the azimuthally averaged temperature PDFs, shown in figure \ref{fig:PDF_Resolution} b, c, and d, for $Ra=2\times10^8$ obtained on different grids. The radial position is $0.2342L$ for the underresolved ($97 \times 49 \times 193$) and $0.2314L$ for the fully resolved ($193 \times 65 \times 257$) simulations. The temperature PDF for the fully resolved simulations averaged over $3000$ dimensionless time units is indicated in black. The green line indicates the result using half of the time series. The temperature PDF averaged over $3000$ dimensionless time units for the underresolved simulations is indicated in blue, and the red indicates the result using half of the time series.  b) Temperature PDF at mid height. c) Temperature PDF at the distance $\lambda_{\theta}^{sl}$ from the bottom plate. d) Temperature PDF at the distance $\lambda_{\theta}^{sl}$ from the top plate.}
\label{fig:PDF_Resolution}
\end{figure}

\begin{figure}
\centering
\includegraphics[width=3.25in]{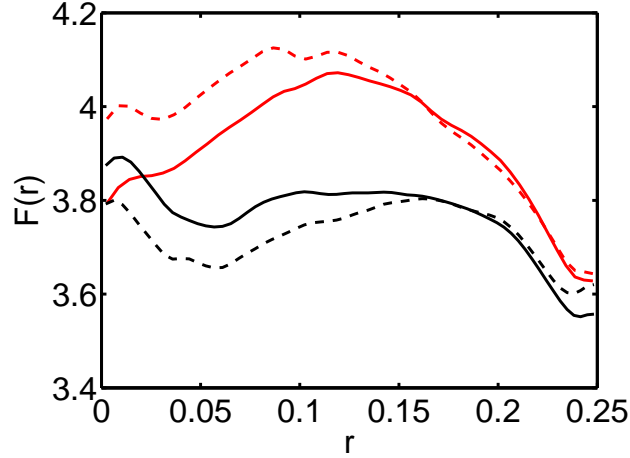}
\caption{Flatness of the temperature PDF at mid height for $Ra=2\times10^8$ for the underresolved (red, $97 \times 49 \times 193$) and the fully resolved simulations (black, $193 \times 65 \times 257$). The solid lines indicate the result after averaging over $3000$ dimensionless time units and the dashed lines the result after averaging over $1600$ dimensionless time units. Both simulations are started from the same initial field obtained at a lower $Ra$ and the data collecting is started when each simulation has reached the statistically stationary state.}
\label{fig:Flatness}
\end{figure}

To further investigate the influence of the grid resolution, we calculated azimuthally averaged PDFs (see also \cite{emr08,kun08,shi07,shi08,kac09}) of the temperature averaged over $3000$ dimensionless time units for $Ra=2\times 10^8$, comparing the underresolved case ($97 \times 49 \times 193$) with the fully resolved one ($193 \times 65 \times 257$). Figure \ref{fig:PDF_Resolution} shows that the temperature PDFs at mid height and at a distance $\lambda_\theta^{sl}$ (thermal BL based on the slope) from the plates have longer tails in the underresolved simulation than in the fully resolved one. Again the reason lies in the rising (falling) plumes from the bottom (top) plate which are not properly dissipated in the underresolved simulations and therefore travel further from the plates. The comparison with the PDF obtained using half of the time series reveals that the differences in the PDFs are not due to a lack of averaging, but due to insufficient grid resolutions. We note that we observe similar differences at other radial positions, only the averaging around the cylinder axis ($r=0$) leads to not fully converged results due to the geometry. In figure \ref{fig:Flatness} we show the effect of the grid resolution on the flatness obtained at mid height for fully resolved and underresolved simulations. Comparison between the solid and dashed lines show that the data are converged close to the sidewall where the statistics is best due to geometric reason. Comparison between black (well resolved) and red (underresolved) reveals that the insufficiently dissipated plumes mainly close to the sidewall leads to too large flatnesses in the underresolved simulations.

Although the bulk is turbulent, scalingwise the BLs still behave in a laminar way due to the small BL Reynolds number (\cite{ahl09}). Therefore we compare the thermal BL profile obtained from the simulations with the Prandtl-Blasius (PB) profile, as done by \cite{sug09} for $2D$ RB simulations. The temperature gradient of the PB profile is matched to the temperature gradient obtained in the high resolution simulation. The temperature profile obtained in the simulations best matches the PB profile around the cylinder axis ($r=0$). Close to the sidewall the agreement is worse due to the rising (falling) plumes in this region. We now compare the difference between the PB profile and the result obtained from the simulation for different $Ra$. We determine, at the cylinder axis, ($\theta_{sim}$-$\theta_{PB}$)/($\Delta$-$\theta_{PB})$ for the bottom BL and ($\theta_{PB}$-$\theta_{sim}$)/($\theta_{PB}$) for the top BL. Here $\theta_{sim}$ is the mean temperature at a distance $\lambda_{\theta}^{sl}$ from the plate and $\theta_{PB}$ the temperature according to PB at this height, after having matched the gradient at the plate to the simulation data. If the simulation exactly matched PB (e.g. for very small $Ra$), this expression would be zero. In contrast, here for $Ra = 2 \times 10^8$ ($2\times 10^9$, $2\times 10^{10}$), it is $0.103$, ($0.130$, $0.149$). As expected, the expression is smaller for the lower $Ra$ numbers. We perform the same procedure for our previous results of \cite{zho09b} at $Ra = 1 \times 10^8$ with $\Gamma=1$ and now different $Pr$. For $Pr=0.7$, $Pr=6.4$, and $Pr=20$ we now obtain $0.099$, $0.040$, and $0.033$, respectively. Now the expression is closest to zero at the higher $Pr$, as then the $Re$ number is lower and PB holds better.

\begin{figure}
\centering
\subfigure{\includegraphics[width=0.49\textwidth]{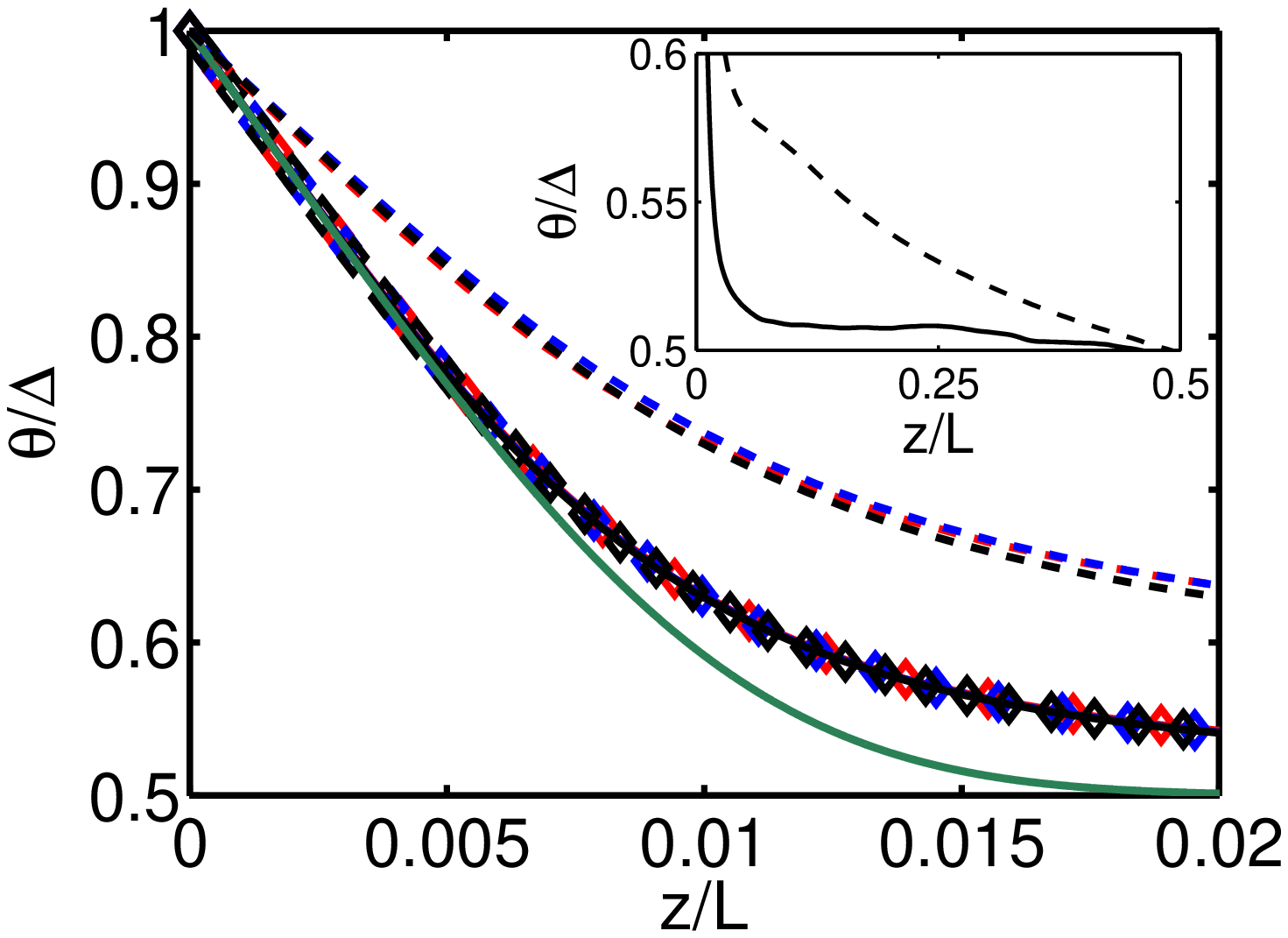}}
\subfigure{\includegraphics[width=0.49\textwidth]{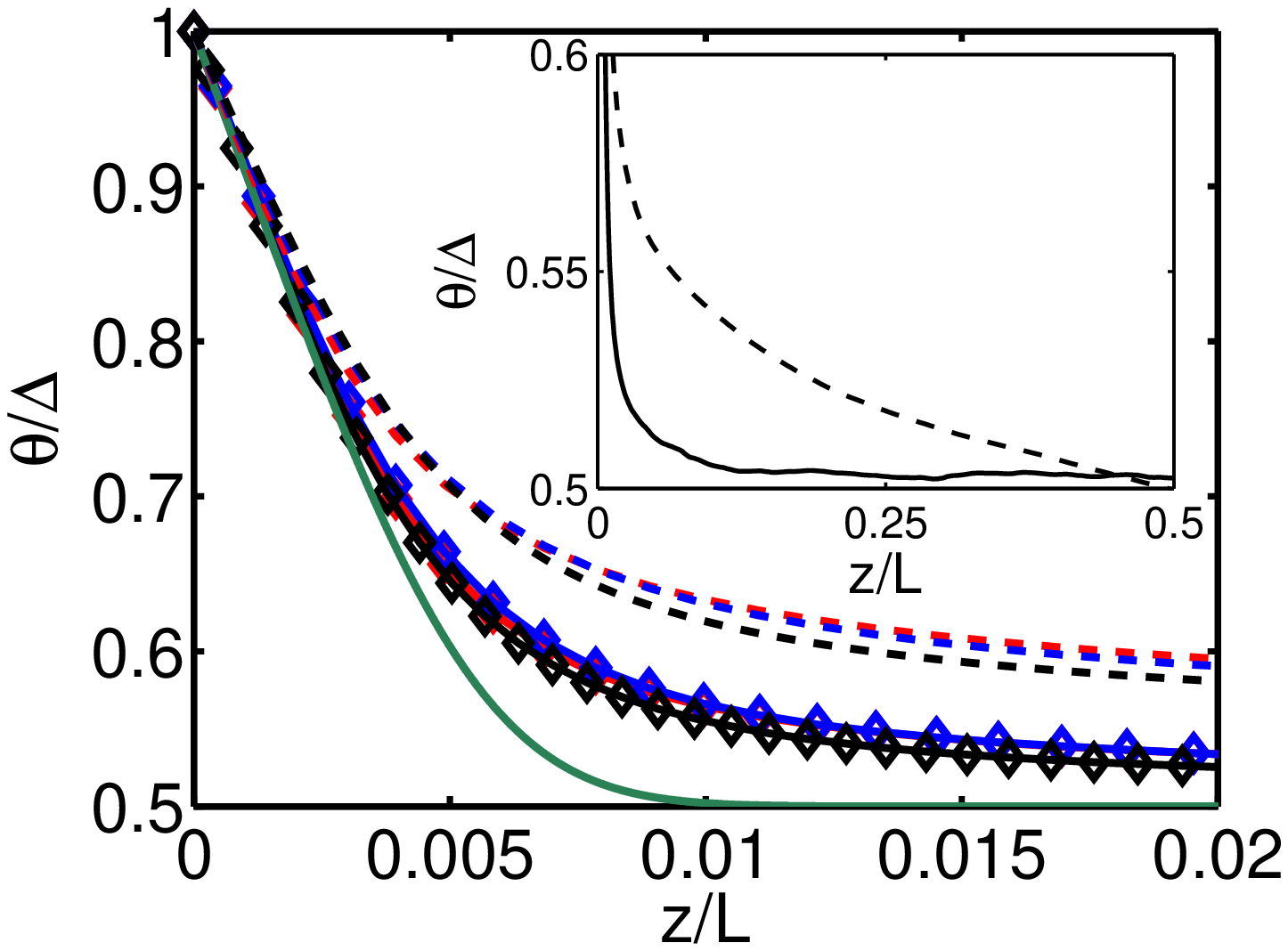}}
\caption{Azimuthally averaged temperature profiles obtained from the simulations at different grids. a) $Ra=2 \times 10^8$, for the grids $97 \times 49 \times 193$ (red), $193 \times 65 \times 257$ (blue), $257 \times 97 \times 385$ (black) and b) $Ra=2 \times 10^9$ for the grids $129 \times 65 \times 257$ (red), $193 \times 65 \times 257$ (blue), $385 \times 97 \times 385$ (black). The diamonds indicate the axial position of the grid points. The solid lines show the temperature profile at the cylinder axis ($r=0$) and the dashed lines at the radial position $0.225L$. The green line indicates the PB profile matched to the temperature gradient at the cylinder axis ($r=0$) of the high resolution simulation. The insets show the temperature profile from the highest resolution data over a larger axial range. Here the solid line indicates the profile at the axis and the dashed line the temperature profile at the radial position $0.225L$.}
\label{fig:Blasius}
\end{figure}

\begin{figure}
\centering
\subfigure[]{\includegraphics[width=0.49\textwidth]{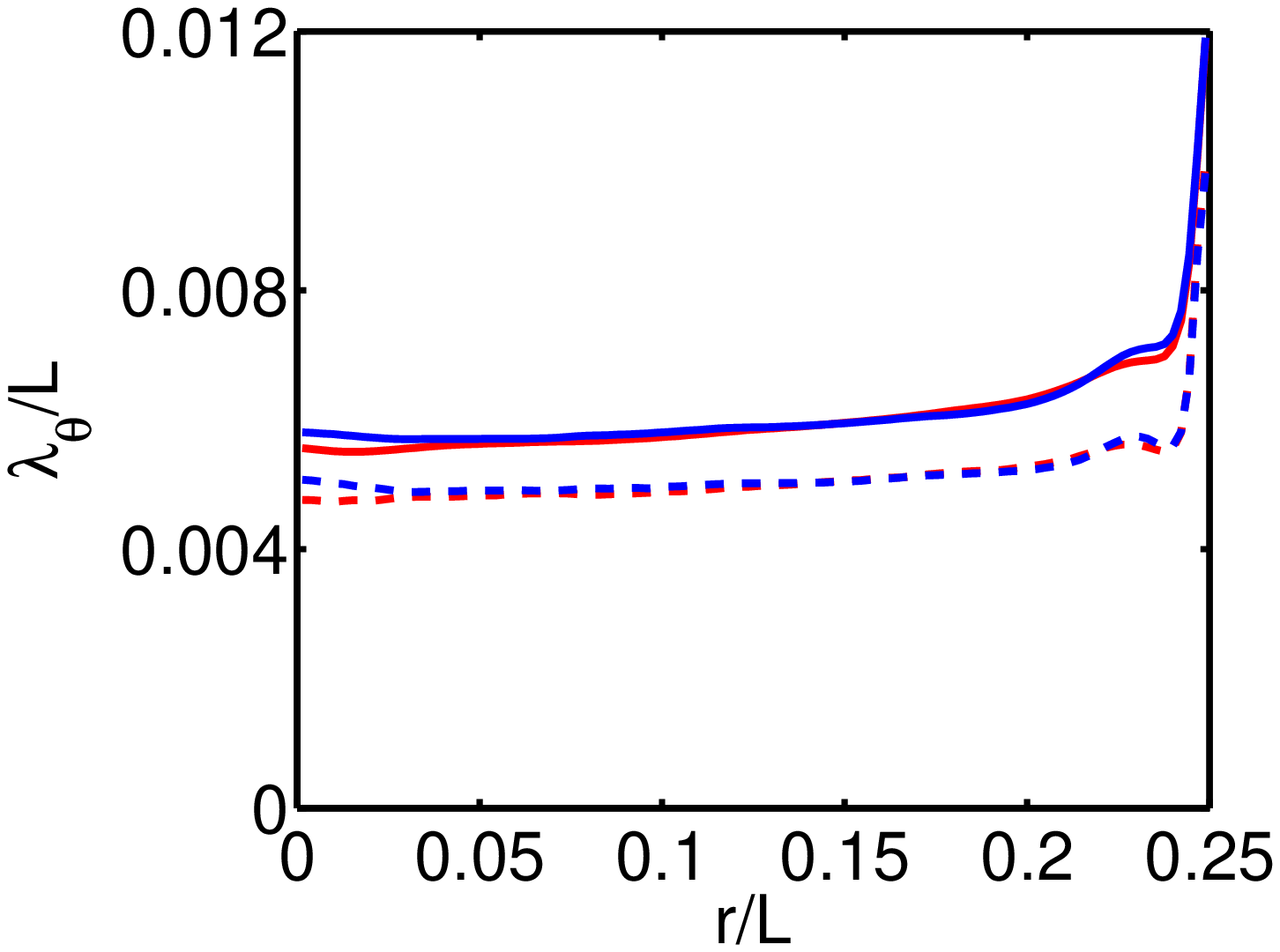}}
\subfigure[]{\includegraphics[width=0.49\textwidth]{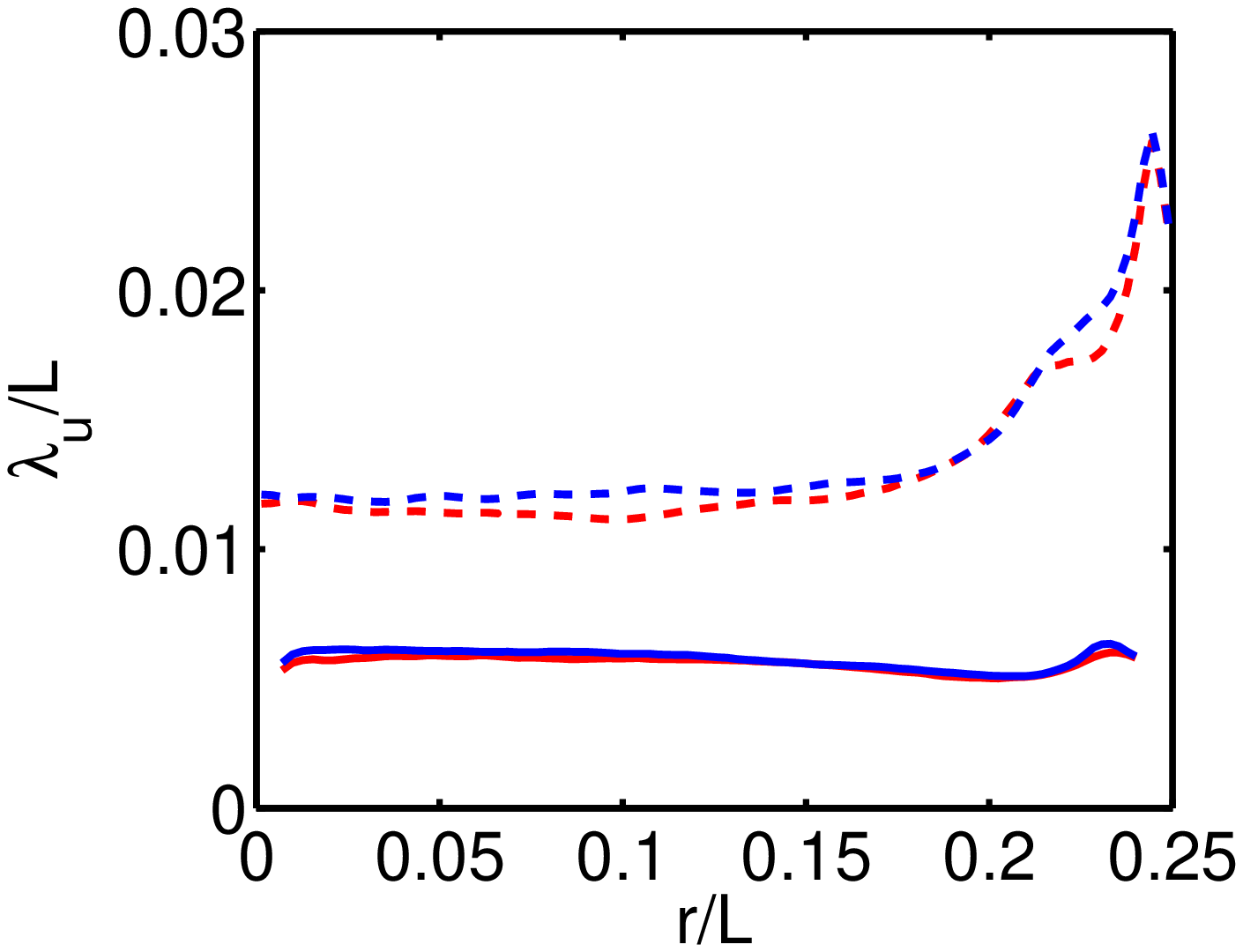}}
\subfigure[]{\includegraphics[width=0.49\textwidth]{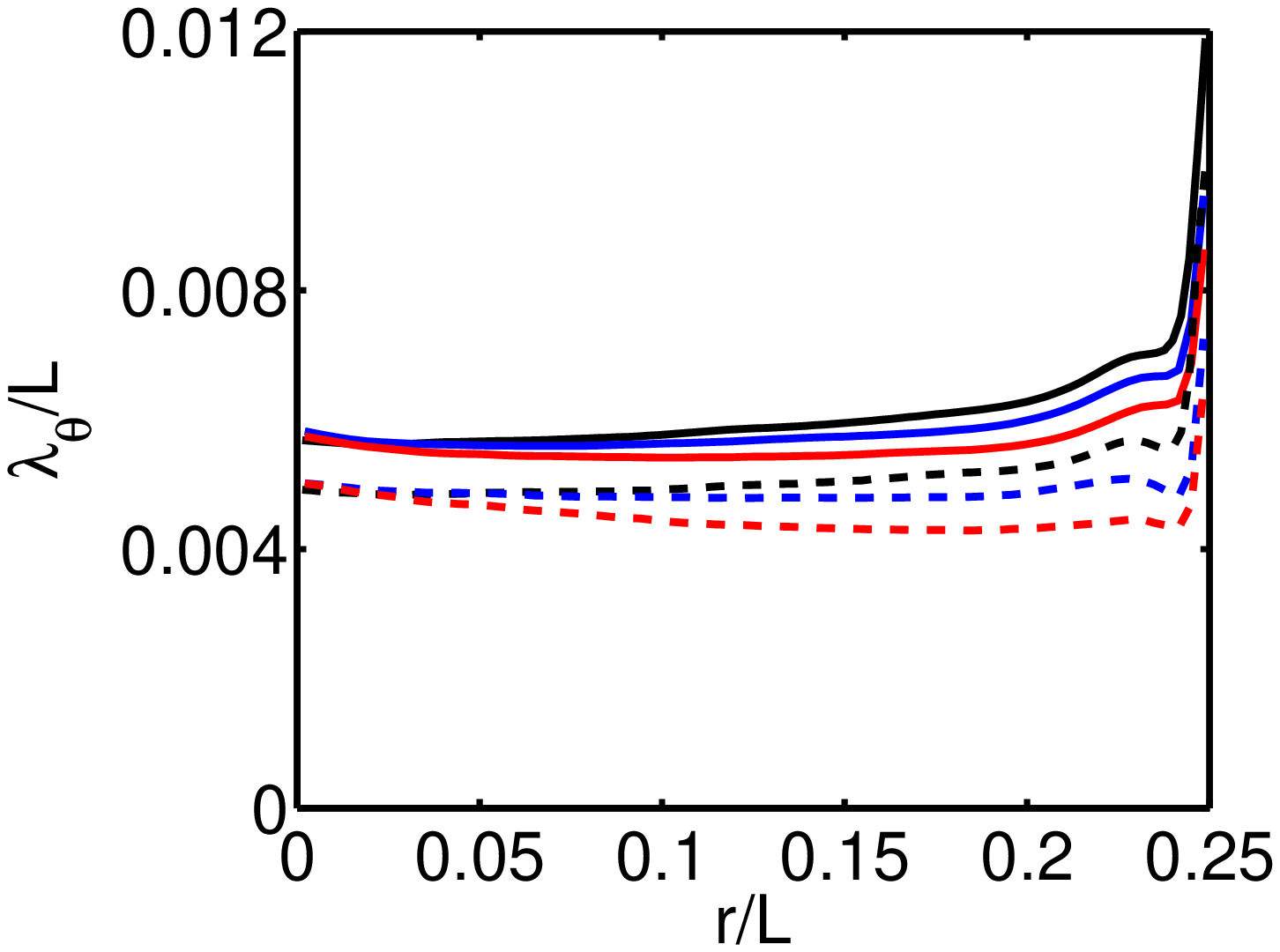}}
\subfigure[]{\includegraphics[width=0.49\textwidth]{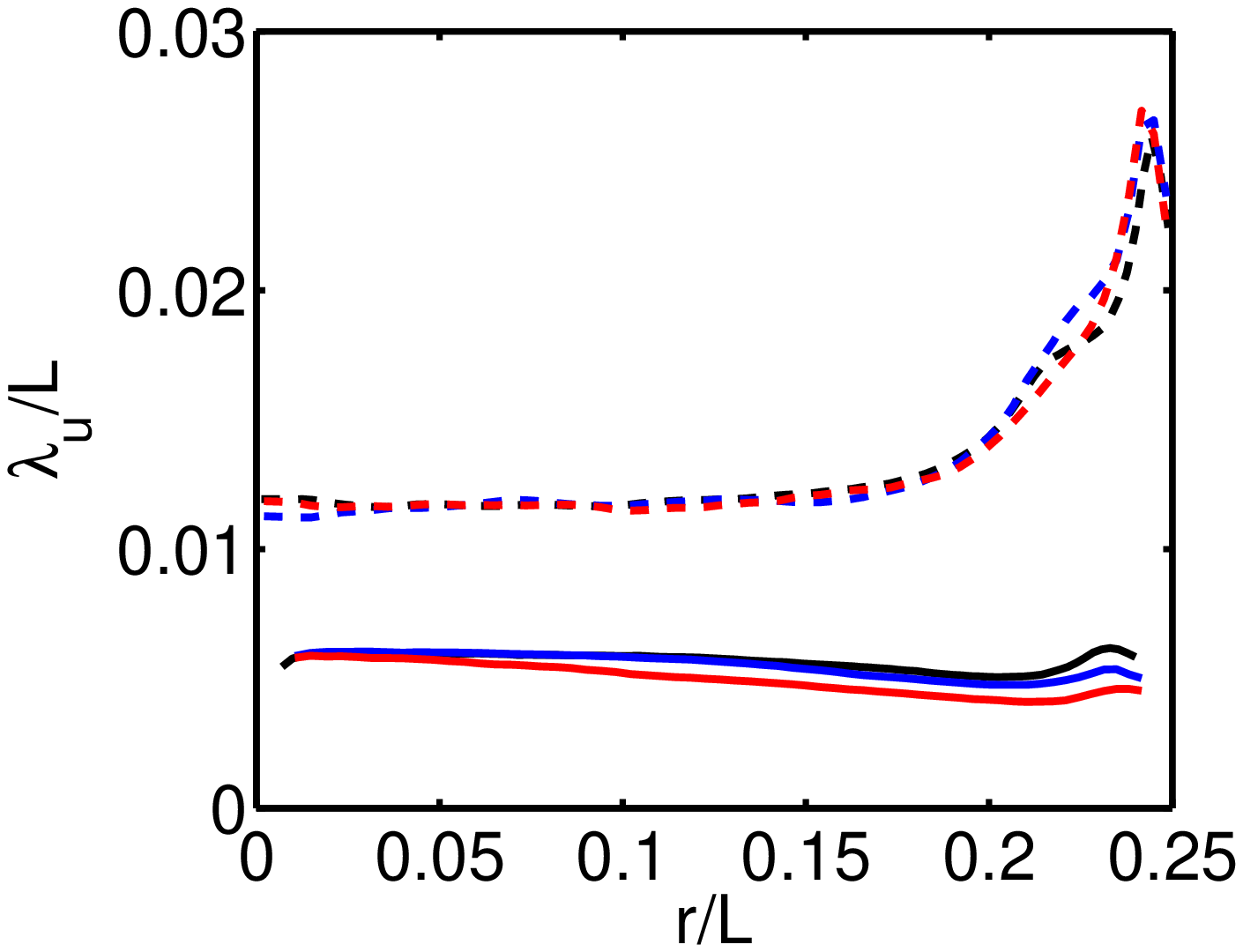}}
\caption{Azimuthally averaged BL thicknesses as function of the radial position for $Ra=2 \times 10^9$. Figure \ref{fig:Radial}a and b show data from the high resolution simulation ($385 \times 97 \times 385$). a) The solid line indicates $\lambda_{\theta}^{sl}$ and the dashed lines $\lambda_{\theta}^{rms}$, where red and blue indicate the bottom and top plate, respectively. b) The solid line indicates $\lambda_{u}^{\epsilon_u}$ and the dashed line $\lambda_{u}^{rms}$ based on the azimuthal velocity (colors as in figure \ref{fig:Radial}a). c) Now the colors indicate the different grid resolutions: Red ($129 \times 65 \times 257$), blue ($193 \times 65 \times 257$), and black ($385 \times 97 \times 385$). The solid lines indicate $\lambda_{\theta}^{sl}$ and the dashed lines $\lambda_{\theta}^{rms}$. The data for the bottom and top BL are averaged for clarity. Note that the BL is thicker (especially close to the sidewall) in the higher resolution simulations, which is in agreement with the observed $Nu$ trend. d) The solid lines indicate $\lambda_u^{\epsilon_u}$ and the dashed lines $\lambda_u^{rms}$ based on the azimuthal velocity for the different grid resolutions (colors as in figure \ref{fig:Radial}c).}
\label{fig:Radial}
\end{figure}

Figure \ref{fig:Radial} shows the radial dependence of $\lambda_{\theta}^{sl}$, $\lambda_{\theta}^{rms}$ (thermal BL thickness based on maximum rms value), $\lambda_{u}^{rms}$ (kinetic BL thickness based on maximum azimuthal rms velocity), and $\lambda_{u}^{\epsilon_u}$ (kinetic BL thickness defined as the axial position of the maximum kinetic energy dissipation rate, multiplied by $2$). $\lambda_{u}^{rms}$ is widely used in literature to define the kinetic BL thickness; however, this definition overestimates the kinetic BL thickness. $\lambda_{u}^{\epsilon_u}$ defines the  BL as the region where the kinetic dissipation is highest and it is this region
 where a particular good resolution is required. Such defined kinetic BL thickness now  well agrees with
that of the thermal BL, $\lambda_u^{\epsilon_u} \approx \lambda_\theta^{sl}$, as  from  PB theory expected for the kinetic BL, once $Pr \sim 1$. Figure \ref{fig:Radial} shows that both BLs become thicker closer to the sidewall. This is due to the plumes traveling along the sidewall and lower velocities very close to the sidewall. {\footnote{
We note that  therefore in that region the definition of $\lambda_{u}^{\epsilon_u}$ misrepresents the BL thickness.}}
Thus the enhanced grid resolution in the vertical direction near the plates is most important around the cylinder axis ($r=0$). In contrast, the azimuthal (and radial) resolution are most important when properly resolving the flow close to the sidewall. Note that the difference in the BL thicknesses between the fully resolved and underresolved simulations is largest close to the sidewall, demonstrating that this is indeed a delicate region from a resolution point of view.

\section{Conclusions}

In summary, the high-resolution results using constant temperature conditions are in good agreement with the experimental data, see figure \ref{fig:Nusselt}. It thus turns out that a good resolution of $\ell_{max} \sim \eta,\eta_T$ is crucial to obtain reliable results for Nusselt number. Close attention has to be given to the resolution used in all directions (azimuthal, radial, and axial). In particular, the azimuthal resolution is crucial to guarantee sufficient plume dissipation close to the sidewall. In underresolved simulations the exact relation $\epsilon_\theta=\kappa \Delta^2 Nu/L^2$ for the thermal dissipation rate does not hold. Hot (cold) plumes travel further from the bottom (top) plate than in the fully resolved simulation, because they are not sufficiently dissipated. We also showed that there is a strong radial dependence of the BL structures. At the cylinder axis (r=0) the temperature profile obtained in the simulations agrees well with the PB case, whereas close to the sidewall the agreement is worse due to rising (falling) plumes in this region.

The effect of changing the constant temperature condition at the bottom plate to a constant heat flux condition will be discussed in detail in a forthcoming publication.

\begin{acknowledgments}
\noindent
{\it Acknowledgment:} We thank S. Grossmann and G. Ahlers for discussions and G.W. Bruggert for drawing figure 1b. The work in Twente was supported by FOM and the National Computing Facilities (NCF), both sponsored by NWO. The simulations up to $Ra=2 \times 10^{10}$ on the $513 \times 129 \times 513$ grid were performed on the Huygens cluster (SARA). The simulation at $2 \times 10^{10}$ on the $385 \times 257 \times 1025$ grid and the $2 \times 10^{11}$ simulations were performed at the computing center CASPUR in Roma. Support from Drs. F. Massaioli and G. Amati is gratefully acknowledged.
\end{acknowledgments}

\bibliographystyle{jfm}

\bibliography{C:/Users/local.la/Biblatex/literatur}

\end{document}